\documentclass[aps,twocolumn,pra,showpacs,floatfix]{revtex4}
\usepackage{epsfig}
\usepackage{graphicx}
\usepackage{dcolumn}
\usepackage{amssymb,amsmath}
\usepackage{mathrsfs}
\usepackage{xcolor}
\begin{document}

\title{Parity non-conservation in the isotope chain of tin}

\author{V. A. Dzuba$^1$, V. V. Flambaum$^1$, D. DeMille$^{2,3}$,  Jianwei Wang$^2$, Geoffrey Zheng$^3$}

\affiliation{$^1$School of Physics, University of New South Wales, Sydney 2052, Australia}

\affiliation{$^2$William H. Miller III Department of Physics and Astronomy, Johns Hopkins University, Baltimore, Maryland 21218 USA}

\affiliation{$^3$James Franck Institute and Department of Physics, University of Chicago, Chicago, Illinois 60637, USA}


\begin{abstract}

 We calculate parity non-conservation (PNC) amplitudes for all magnetic-dipole (M1) transitions within the ground $5p^2$ configuration of Sn, including the standard model interaction and contribution of a hypothetical additional $Z'$-boson.
Among the transitions considered, the $^1$S$_0$-$^3$P$_1$ transition has the largest PNC amplitude and appears to be the most promising candidate for an experiment. We also discuss a measurement method capable of achieving unprecedentedly high precision in a measurement of PNC in this transition. We argue that the most robust test should be based on ratios of PNC amplitudes for different isotopes, since the atomic-structure factor largely cancels in such ratios. We study the effect of the neutron skin on these isotope ratios using available nuclear data for Sn and show that the uncertainty associated with the neutron skin can be reduced to the 
 $10^{-4}$ level relative to the isotopic variation of the PNC effect. Our results indicate that PNC measurements along a chain of Sn isotopes offer a realistic and sensitive probe of new physics.

\end{abstract}


\maketitle

\section{introduction}

The study of parity non-conservation (PNC) in atoms is a powerful tool for testing the Standard Model (SM) at low energies and for searching for new physics beyond it. The most accurate atomic PNC measurements have been performed for Cs (with an accuracy of 0.35\%~\cite{Wood}). When combined with state-of-the-art calculations (with an accuracy better than 0.5 \% ~\cite{DzuFlaSus89a,BluJohSap1990,DzuFlaGin02,CsPNC12,DereviankoCsPNC}), this has enabled the extraction of the weak nuclear charge of $^{133}$Cs, which is in good agreement with the SM~\cite{SM,Samsonov}.

Further progress in this direction is challenging, primarily due to limitations in atomic-structure calculations. An alternative approach, suggested in Ref.~\cite{DzuFlaKhr86}, is to measure ratios of PNC amplitudes between different isotopes of the same element. In such ratios, the electronic-structure factor obtained from atomic calculations largely cancels out, making the result essentially independent of these calculations. At the same time, the ratio contains valuable information; in particular, it is sensitive to possible manifestations of new physics, e.g. new contact interactions which affect nuclear weak charge ~\cite{DzuFlaKhr86,Brown,Viatkina}.
Measurements of the ratios of  PNC effects for  different isotopes of Yb atoms have been performed in Refs. ~\cite{Antypas2018,Antypas2019}, putting constraints on electron-proton and electron-neutron interactions mediated by a light $Z'$ boson 
(see also the proposal ~\cite{Orozco} for Fr).

The PNC ratio is sensitive to the neutron distribution in the nucleus, whose radius is typically larger than that of the proton distribution. This difference is referred to as the neutron skin.  The proton distribution radius is usually known to a high accuracy form electron scattering experiments. Neutrons have no electric charge and their  distribution is harder to measure. 

The neutron skin varies from isotope to isotope, affecting PNC amplitudes and potentially mimicking the effects of new physics.
Sensitivity to the neutron skin, investigated in Refs.~\cite{Fortson,Brown,Viatkina}, is proportional to $(Z\alpha)^2$ and is suppressed in light elements.
It was also concluded that the neutron-skin contribution can be estimated with sufficient accuracy to distinguish it from possible manifestations of new physics.
  A first experimental implementation, using the Yb isotope chain,
  was published in Refs.~\cite{Antypas2018,Antypas2019}. 
A combination of single-isotope and isotope chain methods allows
one to measure separately both proton and neutron weak
charges.

In the present work, we consider PNC measurements for a chain of tin isotopes (Sn, $Z=50$). There are several reasons for this choice. First, tin has 10 stable isotopes, including seven with zero nuclear spin ($I=0$) and three with $I=1/2$. Second, sufficiently accurate data on neutron skin are available for many Sn isotopes~\cite{Trzcinska,Terashima,Roca-Maza,Tagami}. The data implies that the effect of neutron skin is relatively small. Finally, Sn is lighter than all atoms in which PNC effects have already been measured. This implies a smaller neutron skin sensitivity ( which is proportional to  $Z^2 \alpha^2$), where $Z$ is the nuclear charge and $\alpha$ is the fine structure constant.

It also appears possible to obtain unprecedented precision in measurements of PNC by using Sn atoms. Ref.~\cite{Fortson-Ion} pointed out that the relative statistical uncertainty in an optimized PNC measurement is proportional to the linewidth of the transition used.  In Sn, the measurement can be performed using a narrow ``clock'' transition between a pair of levels in the ground electronic configuration. It was recently pointed out that laser cooling and trapping of Sn \cite{SnSimsPaper} is viable, which makes it possible to take full advantage of the narrow natural linewidth---much as in optical atomic clocks. Together, this means that ultra-precise measurements of PNC should be possible in many isotopes of Sn.

Another possible advantage of Sn relative to heavy atoms is  a larger relative contribution from a hypothetical light $Z'$ boson.
Indeed, SM  parity violation effects increase with the nuclear charge as $ Z^2 R(Z \alpha)$ while PNC effects due to the low mass $Z'$ boson have no    such $Z$ dependence.  The relativistic factor is $R(Z \alpha) \approx 1$ for small $ Z \alpha$ and reaches $\sim 10$ in heavy atoms.   Therefore, light atoms  have strong relative enhancement of the ratio of $Z'$ to $Z$ boson contributions,  defined by the factor $ \sim 1/[Z^2 R(Z \alpha)]$, which should allow more accurate separation of this new physics effect.

In this work, we calculate $E1^{\rm PNC}$ and M1 amplitudes, as well as their ratios, for all transitions between states of the ground configuration of Sn. Similar calculations for Pb, where both experimental and earlier theoretical data on PNC are available, are used to benchmark the accuracy of our approach. The effect of the neutron skin on PNC in Sn is studied using available experimental data. We find that the contribution of the neutron skin to the ratio of PNC amplitudes is below $10^{-3}$. If higher experimental accuracy is achieved, then, depending on the choice of isotope pair, this effect can either be neglected or separated from possible contributions of new physics.

\section{Method of calculations}

\subsection{Energies and wave functions}
\label{s:EL}
The tin atom has the [Pd]$5s^25p^2$ ground-state electron configuration. The PNC amplitude is dominated by the $5s_{1/2} - 5p_{1/2}$ matrix element of the weak interaction. Therefore, the $5s$ and $5p$ states should be treated on an equal footing as valence states.

This leads us to the $V^{N-4}$ approximation~\cite{Dzu05}, in which the initial Hartree–Fock (HF) calculations are performed for the Pd-like closed-shell Sn~V ion. The basis states for the valence electrons are constructed in the field of the frozen core using the B-spline technique~\cite{B-splines}. The four-electron valence states are then obtained using the configuration interaction (CI) method.

To include core–valence correlations, we combine the CI method with the linearized coupled-cluster single–double (SD) approach~\cite{Dzu-CI-SD14}. Solving the SD equations for both the core and valence states yields two all-order correlation operators acting in the valence space. The operator $\hat{\Sigma}_1$ describes the correlation interaction of a given valence electron with the core, while $\hat{\Sigma}_2$ represents a correction to the Coulomb interaction between two valence electrons induced by core excitations.

These operators are added to the CI Hamiltonian, leading to a significant improvement in the accuracy of the results.

The Breit interaction (magnetic and retardation terms) is also included~\cite{DzuFlaSaf06}, while quantum electrodynamic (QED) corrections are neglected. A proper inclusion of QED corrections to a singular operator, such as the weak interaction, is a complicated task, while the resulting contribution is relatively small (less than 1\%).

To solve the CI problem, we construct a basis of Slater determinants by selecting  two (for even states) or  three (for odd states)  reference configurations and allowing single and double excitations of electrons to virtual orbitals with principal quantum number $n \leq 20$  (in a box of radius  40 $a_B$) and orbital angular momentum $l \leq 4$. The use of more than one reference configuration effectively incorporates selected triple excitations with respect to the dominant configuration. The reference configurations are chosen to represent the leading configurations of the low-lying states of a given parity. In most cases, they can be identified from the spectroscopic classifications of the corresponding states listed in the NIST Atomic Spectra Database~\cite{NIST}. The resulting CI basis is generally sufficient to achieve convergence of the calculated energy levels and other atomic properties.

It is well known that the size of the CI matrix grows rapidly with the number of electrons; for four electrons, it can reach sizes of order $\sim 10^6$. This makes the calculations very challenging. To achieve a reasonable compromise between computational complexity and accuracy, we employ the CIPT (configuration interaction with perturbation theory) method~\cite{CIPT}, which reduces the size of the CI matrix by orders of magnitude with only a minor loss of accuracy.

In this approach, all four-electron basis states $n$ are ordered according to their energies $\langle n|\hat{H}^{\rm CI}|n\rangle$, from low to high. The list is then divided into two unequal subsets: a relatively small low-energy subset and a much larger high-energy subset. The off-diagonal matrix elements between high-energy states are neglected. As a result, the contributions of the high-energy states are treated perturbatively as corrections to the matrix elements within the low-energy subset. This reduces the CI matrix to the size of the low-energy basis (see Ref.~\cite{CIPT} for details). The size of the low-energy subset is a free parameter of the method and is chosen to optimize the balance between accuracy and computational cost.

We found that good balance is achieved when the size of the low-energy basis is $\sim 10^4$, i.e. two orders of magnitude smaller than the total size of the basis. The results for the energy levels of Sn and its heavier analog Pb are presented in sections \ref{s:Sn} and \ref{s:Pb}.


\subsection{M1 amplitudes and  Land\'{e} $g$-factors} 

{For each level, we calculate the Land\'{e} $g$-factor and compare it to both the experimental values and the non-relativistic expression
\begin{equation}\label{e:gf}
g = 1 + \frac{J(J+1)-L(L+1)+S(S+1)}{2J(J+1)}.
\end{equation}
Here, $J$ is the total angular momentum of all atomic electrons, $L$ is the angular momentum, and $S$ is the spin.
The calculated $g$-factor is obtained as the expectation value of the operator describing the interaction between the electrons and the magnetic field (M1): 
$\hat H_{M1} = \overrightarrow{\mu}~\cdot~\overrightarrow{B}$, where $\overrightarrow{\mu}$ is the  pseudo-vector of atomic magnetic moment. In the relativistic case the matrix elements of this operator over single-electron wave functions
\begin {equation}
\psi(r)=\frac{1}{r}\left(\begin{array}{c} f(r)\Omega_{\kappa m} \\ i\alpha g(r) \Omega_{-\kappa m} \end{array} \right)
\end{equation}
are given by 
\begin{eqnarray}
\langle \psi_a|\hat H_{M1}|\psi_b\rangle &=& (\kappa_a+\kappa_b)\langle -\kappa_a||C^1||\kappa_b \rangle \nonumber \\
&\times&\int 2(f_ag_b+g_af_b)J_1(kr)dr B,
\end{eqnarray}
where $\kappa = l$ for $j=l-1/2$, $\kappa = -l-1$ for $j=l+1/2$, $C^1$ is the normalized spherical harmonic, wave vector $k=\omega/c$, 
$J_1(kr)$ is the spherical Bessel function.
The $g$-factor is defined via energy shift $\Delta E = \mu_B g M_J B$, where $M_J$ is the projection of the total atomic angular momentum $J$.
For the many-electron wave function $v$ for valence electrons, we have
\begin{equation}\label{e:gv}
g_v = \frac{1}{\sqrt{J_v(2J_v+1)(J_v+1)}} \langle v|| \hat H_{M1}||v\rangle/B.
\end{equation}
}

\subsection{PNC amplitudes}

The operator of the parity non-conserving nuclear-spin-independent weak interaction mediated by the $Z$-boson can be written as
\begin{equation} \label{e:hpnc}
  \hat W = - \frac{G_F}{2\sqrt{2}} Q_W \gamma_5 \rho(r),
\end{equation}
where $G_F$ is the Fermi constant of the weak interaction, $Q_W$ is the weak nuclear charge, $\gamma_5$ is a Dirac matrix, and $\rho(r)$ is the nuclear density normalised by $\int \rho(r)dV=1$.
We use the Fermi distribution function for $\rho(r)$.
The weak nuclear charge is given by \cite{SM} 
\begin{equation}\label{e:qw}
Q_W = -0.98897N + 0.0706Z,
\end{equation}
where $N$ is the number of neutrons and $Z$ is the number of protons.
New corrections to the weak nuclear charge caused by exchange of two neutrinos, two leptons or two quarks were recently found in Ref.~\cite{Samsonov}.
With this correction included new expression for $Q_W$ reads 
\begin{equation}\label{e:qwn}
Q_W = -0.98207N + 0.071918Z,
\end{equation}

We also consider the exchange of a vector boson $Z'$ 
between electrons and nucleons, which can be described by the following Lagrangian:
\begin{equation}
    L_\text{int} = Z_{\mu}' \sum_{f=e,p,n} \bar{f} \gamma^\mu \left( g_f^V + \gamma_5 g_f^A \right)f.\label{e:Lint}
\end{equation}
The resulting  P-violating interaction between electron axial and nucleon vector currents is described by a Yukawa-type potential:
\begin{equation}
    V_{eN}(r) = \frac{g_e^A g_N^V}{4\pi} \frac{e^{-m_{Z'} r}}{r} \gamma_5,
    \label{e:V12}
\end{equation}
where $r$ is the distance between electron and nucleon, and the $\gamma_5$ matrix corresponds to electron.
 We introduce the shorthand notation $g_N^V = (N g_n^V + Z g_p^V)/A$, where $A= Z + N $ is the nucleon number.

The PNC amplitude between many-electron states $a$ and $b$ is given by
\begin{equation}\label{e:amp}
E1^{\rm PNC}_{ab} = \sum_n \frac{\langle a|\hat W |n \rangle \langle n|\hat D |b \rangle}{E_a - E_n} 
                               + \sum_n \frac{\langle a|\hat D |n \rangle \langle n|\hat W |b \rangle}{E_b - E_n} 
\end{equation}
Here $\hat W$ is the operator of weak interaction (\ref{e:hpnc}), $\hat D$ is the electric dipole operator.
In the first term $J_n = J_a$, in the second term $J_n = J_b$.

We use the Dalgarno and Lewis method to perform the calculations \cite{dalgarno1955exact}. We introduce a correction to the atomic wave functions $|a\rangle$ and $|b\rangle$ caused by an external field $\hat W$,
\begin{equation}\label{e:da}
|\delta a\rangle = \sum_n |n\rangle \frac{\langle n||\hat W ||a\rangle}{E_a-E_n}.
\end{equation}
The correction can be found by solving the equation
\begin{equation}\label{e:dfa}
(H-E_a)|\delta a\rangle = - \hat W |a\rangle.
\end{equation}
Similar equations can be written for state $b$.
When the corrections are found, the amplitude (\ref{e:amp}) is reduced to
\begin{equation}\label{e:DW}
E1^{\rm PNC}_{ab}  = \langle \delta a|| \hat D || b \rangle +  \langle  a|| \hat D || \delta b \rangle.
\end{equation}

\subsection{The RPA method}

Equation (\ref{e:amp}) is accurate when $a$, $n$ and $b$ are many-electron states which include all atomic electrons. 
In practice, the many-body calculations are done for a few valence electrons above closed-shell core. The effect of core electrons on the matrix elements between states of valence electrons can be reduced to the redefinition of the operator of the external field which causes the transition.
This is done in the framework of the random-phase approximation (RPA) method, (see e.g.~\cite{DzuFlaSilSus87}).
The RPA equation for the core have the form
\begin{equation}\label{e:RPA}
(H^{\rm HF} - \epsilon_c)\delta \psi_c = -(F+\delta V_{\rm core}^{F}),
\end{equation}
where $H^{\rm HF}$ is the Hartree-Fock (HF) Hamiltonian, index $c$ numerates states in the core, $\delta \psi_c$ is the correction to the core function due to the effect of external field $F$ (weak $\hat W$, electric dipole $\hat D$, { or magnetic dipole $\hat M$} interactions) and $\delta V_{\rm core}^{F}$ is the correction to the core potential due to the change in all core wave functions. The HF and RPA calculations are done with the same Hamiltonian $H^{\rm HF}$ (i.e., the same core). 
The RPA equations are solved iteratively until self-consistency is achieved. This corresponds to an all-order treatment of the response of the atomic core to the external field within the Coulomb interaction. The resulting modification of the effective interaction is known as the core-polarization (CP) effect.

Then the operator of external field acting in valence space is modified by $ F \rightarrow F+\delta V_{\rm core}^{F}$.
The term with $\delta V_{\rm core}^{F}$ is called the core polarization correction.

Core polarization and correlations provide the most important corrections to matrix elements of external fields. The CP correction can be reduced to a redefinition of the operator, while correlation effects are incorporated into the wave functions.

There are also several smaller corrections, such as structure radiation and wave-function renormalization. These are neglected in the present work, as their contributions are smaller than the uncertainty associated with the incomplete treatment of correlations.

\subsection{The PNC amplitude in the CIPT calculation}

Calculations of PNC amplitudes are very similar to the calculations of polarizabilities~\cite{Symmetry}.
First, we need to find the many-electron states $a$ and $b$. This is done by solving the matrix eigenvalue problem with the effective CI matrix~\cite{CIPT},
\begin{eqnarray}
&& H_a^{\rm CI}X_a = E_a X_a \\
&& H_b^{\rm CI}X_b = E_b X_b. 
\end{eqnarray}
Note that the effective CI matrix $H^{\rm CI}$ depends on parity and the value of the total angular momentum of the atom $J_a$ or $J_b$.
Since $J_a$ and $J_b$ can be different, the CI matrix has an index $a$ or $b$. Vectors $X_a$ and $X_b$ are sets of expansion coefficients over four-electron basis states which have definite parity and $J$. These states are constructed from four-electron single-determinant states of a specific configuration by diagonalising the matrix of the $J^2$ operator.
To find corrections $\delta a$ and $\delta b$ to the valence wave functions caused by weak interaction we rewrite Eq.~(\ref{e:dfa}) in a matrix
form
\begin{eqnarray}
&& \left(H_{a'}^{\rm CI} - E_a\right)Y_a = - W X_a \\
&& \left(H_{b'}^{\rm CI} - E_b\right)Y_b = - W X_b. 
\end{eqnarray}
Here $W$ is an operator of the weak interaction. It is a sum over valence electrons which includes the core polarisation correction (see previous subsection), $W = \sum_{i=1}^4 (\hat W + \delta V^{W})_i$. 
We use indices $a'$ and $b'$ for the effective CI matrices to stress that states $a$, $b$ and their corrections due to the weak interaction have different parity. 

 When the vectors $X_a$, $X_b$, $Y_a$, $Y_b$ are found, the PNC amplitude (\ref{e:amp}) is given by
 \begin{equation}\label{e:ampm}
 E1^{\rm PNC}_{ab} = Y_a D X_b + X_a D Y_b.
 \end{equation}
Here again $D$ is an electric dipole operator which includes the core polarisation correction, $D = \sum_{i=1}^4 (\hat D + \delta V^{D})_i$. 

We stress once more that calculation of PNC amplitudes is very similar to the calculation of polarizabilities~\cite{Symmetry}. Therefore, we calculate
polarizabilities too to have another test of the accuracy of calculations.

\section{Results}

\subsection{PNC in Pb}

The Pb atom provides a convenient testing ground for assessing the accuracy of the calculations and for fine-tuning the computational parameters (e.g., the size of the low-energy basis; see Sec.~\ref{s:EL}). The Pb atom has an electron configuration similar to that of Sn, and therefore the accuracy of the calculations is expected to be comparable.

The PNC effect in the transition between the ground and first excited states of Pb has been studied both theoretically~\cite{DzuFlaSilSus88,Safronova} and experimentally~\cite{Pb-PNC1,Pb-PNC2,Pb-PNC3} by several groups.

The calculations begin with the determination of energy levels. The calculated energies and $g$-factors for the lowest states of Pb are compared with the NIST data in Table~\ref{t:ELPb}. The agreement for the energies is at the level of about 1\%, indicating that the accuracy of the wave functions is also expected to be good.

\label{s:Pb}

\begin{table}
 \caption{\label{t:ELPb}
 Energy levels ($E$) and $g$-factors of lowest states of Pb.}
\begin{ruledtabular}
\begin{tabular}{lll c rcrc}
&&&\multicolumn{2}{c}{NIST\cite{NIST}}&
\multicolumn{2}{c}{Calculations}\\
\multicolumn{2}{c}{State}&
\multicolumn{1}{c}{$J$}&
\multicolumn{1}{c}{$E$}&
\multicolumn{1}{c}{$g$}&
\multicolumn{1}{c}{$E$}&
\multicolumn{1}{c}{$g$}\\
&&&\multicolumn{1}{c}{[cm$^{-1}$]}&&
\multicolumn{1}{c}{[cm$^{-1}$]}\\
\hline
$6s^26p^2$ & (1/2,1/2) &  0 &      0.000  &    0    &    0    &  0 \\
           &	       &	  &	        &	  &	   & \\
$6s^26p^2$ & (3/2,1/2) &  1 &   7819.2626 &   1.501 &   7820 &   1.500 \\ 
           &           &  2 &  10650.3271 &   1.269 &  10827 &   1.279 \\
           &	       &	  &	        &	  &	   & \\
$6s^26p^2$ & (3/2,3/2) &  2 &  21457.7982 &   1.230 &  21641 &   1.221 \\
           &           &  0 &  29466.8303 &    0    &  29137 & 0 \\
           &	       &	  &	        &	  &	   & \\
$6s^26p7s$ & (1/2,1/2)$^o$ & 0 &  34959.9084 &    0    &  35307 & 0 \\
           &           &  1 &  35287.2244 &   1.349 &  35662 &  1.350 \\
\end{tabular}
\end{ruledtabular}
\end{table}

Table~\ref{t:PbPNC} presents the PNC-related parameters of Pb calculated in the present work, along with a comparison to earlier calculations and measurements. This comparison allows us to assess the accuracy of the present calculations.

The calculated value of the static dipole polarizability of the ground state, $\alpha_0$, is about 7\% lower than the recommended value from Ref.~\cite{PolTab}. This deviation is comparable to the uncertainty of the recommended value (6\%).
The calculations follow the procedure developed in Ref.~\cite{Symmetry}.

The effect of PNC in the transition between ground and first excited states of Pb was measured in several experimental works~\cite{Pb-PNC1,Pb-PNC2,Pb-PNC3}.
The measured value is the ratio ($R$) of the $E1^{\rm PNC}$ to the M1 amplitudes,
\begin{equation}\label{e:R}
R = \frac{{\rm Im}(E1^{\rm PNC})}{M1}.
\end{equation}
Therefore, we calculate all three parameters. For ease of comparison with earlier work, the PNC amplitudes are presented both as reduced matrix elements (RMEs) and as $z$-components, while the M1 amplitudes are given in terms of RMEs only. 
The values of $E1^{\mathrm{PNC}}$ and $R$ from Ref.~\cite{Safronova} were multiplied by the factor $(-Q_W/N)$ to convert them to the same units as the other entries in the table, including the experimental results.


The agreement of the present calculations with the experimental PNC data is very good (at the level of $\sim 4\%$). The agreement with  the  calculations of Ref.~\cite{Safronova}  and our previous calculation Ref.~\cite{DzuFlaSilSus88} is also good. 

Note that the measurements  ~\cite{Pb-PNC1,Pb-PNC2,Pb-PNC3} were performed after our calculation Ref.~\cite{DzuFlaSilSus88}, so excellent agreement indicates reliability of our predictions. However, calculations in the presented work have been performed by a different method.  

We therefore conclude that the accuracy of the present calculations of the PNC amplitudes is at the level of approximately 4--5\%. A similar level of accuracy is expected for Sn. 
This is significantly lower than the $\sim 0.5\%$ accuracy achieved for the PNC amplitude in Cs~\cite{DzuFlaSus89a,BluJohSap1990,DzuFlaGin02,CsPNC12,DereviankoCsPNC}. The main limitation arises from the incomplete treatment of correlations among the valence electrons. This problem is absent in Cs and other alkali-metal atoms, which have only a single valence electron.

In general, the accuracy of atomic-structure calculations decreases as the number of valence electrons increases. As discussed in the Introduction, the principal motivation for considering Sn in PNC studies is its large number of stable isotopes. Measurements of the ratios of PNC amplitudes for different isotopes can provide valuable information on neutron-skin effects and possible manifestations of physics beyond the Standard Model. Moreover, the interpretation of these ratios practically does not  dependent on the accuracy of the atomic-structure calculations because  theoretical uncertainty cancels in the ratio.

\begin{table}
 \caption{\label{t:PbPNC}  PNC-related parameters of $^{207}$Pb; $\alpha_0$ is static dipole polarizability of the ground state, 
 M1 and $E1^{\rm PNC}$ amplitudes refer to the transition between ground and first excited states; RME and $z$-component of the PNC amplitude are in units of $10^{-10}  iea_B$.
}
\begin{ruledtabular}
\begin{tabular}{lll}
\multicolumn{1}{c}{Parameter}&
\multicolumn{1}{c}{This}&
\multicolumn{1}{c}{Other}\\
\multicolumn{1}{c}{and units}&
\multicolumn{1}{c}{work}& \\
\hline
$\alpha_0$ [$a_B^3$]& 43.5 & 46.5~\cite{Safronova}; 47(3)\footnotemark[1]~\cite{PolTab} \\
M1(RME) [$\mu_B$] & 1.293 & 1.283~\cite{DzuFlaSilSus88}; 1.293~\cite{Safronova} \\
$E1^{\rm PNC}$(RME) & -4.86 & -4.6~\cite{DzuFlaSilSus88}; -4.72~\cite{Safronova} \\
$E1^{\rm PNC}_z$ & 2.80 & 2.63~\cite{DzuFlaSilSus88} \\ 
$R$ [$10^{-8}$] & -10.3 & -9.8(8)~\cite{DzuFlaSilSus88}; -10.4(8)~\cite{Safronova}; \\
&&-9.86(12)\footnotemark[2]~\cite{Pb-PNC1,Pb-PNC2}; -9.80(33)\footnotemark[2]~\cite{Pb-PNC3} \\
\end{tabular}
\footnotetext[1]{Recommended value based on a number of experimental and theoretical values.}
\footnotetext[2]{Experimental values.}
\end{ruledtabular}
\end{table}

\subsection{PNC in Sn}

\label{s:Sn}

The calculated energies and $g$-factors for the lowest even- and odd-parity states of Sn are compared with the NIST data in Table~\ref{t:EL}. For the even states of the ground configuration, we also present lifetimes and static dipole polarizabilities.

The agreement for the energies of the odd states is at the level of about 1\%, similar to the case of Pb. For the low-lying even states, the agreement is slightly worse. This can be explained by the subtle interplay between relativistic and correlation effects. Relativistic effects shift the $s$ and $p_{1/2}$ states downward in energy (i.e., closer to the core), while the other states are shifted upward. This leads, in particular, to a larger binding energy of the valence electrons.

The four-electron removal energy for the ground state of Pb is about 9\% larger than that of Sn. Another manifestation of strong relativistic effects in Pb is the high excitation energy of the $6s$ electron. The NIST data~\cite{NIST} do not include any states with excitations from the $6s$ subshell in the discrete spectrum. In contrast, states involving excitation from the $5s$ subshell appear at $E \sim 40000~\mathrm{cm}^{-1}$ in Sn (see Table~\ref{t:EL}). The lower excitation energy facilitates stronger configuration mixing, which in turn reduces the accuracy of the calculations.
Note however that the difference between theory and experiment is about 0.03\%  of the four-electron removal energy.

Lifetimes of even states are calculated via M1 transitions to all lower states. Electric quadrupole (E2) transitions give negligible contributions and are not included.

Table ~\ref{t:EL} presents calculated polarizabilities for all even states of the ground configuration. 
The value for the ground state is  in good agreement with the recommended value from Ref.~\cite{PolTab}, similar to what takes place for Pb (see Table~\ref{t:PbPNC}). 

\begin{table}
 \caption{\label{t:EL}
 Energy levels ($E$), $g$-factors, lifetimes ($\tau$) and static scalar polarizabilities ($\alpha_0$) of lowest states of Sn.}
\begin{ruledtabular}
\begin{tabular}{lll rcrc cl}
&&&\multicolumn{2}{c}{NIST\cite{NIST}}&
\multicolumn{4}{c}{Calculations}\\
\multicolumn{1}{c}{$N$}&
\multicolumn{2}{c}{State}&
\multicolumn{1}{c}{$E$}&
\multicolumn{1}{c}{$g$}&
\multicolumn{1}{c}{$E$}&
\multicolumn{1}{c}{$g$}&
\multicolumn{1}{c}{$\tau$}&
\multicolumn{1}{c}{$\alpha_0$}\\
&&&\multicolumn{1}{c}{[cm$^{-1}$]}&&
\multicolumn{1}{c}{[cm$^{-1}$]}&&
\multicolumn{1}{c}{[$s$]}&
\multicolumn{1}{c}{[$a_0^{3}$]}\\
\hline
 1 & $5s^25p^2$ & $^3$P$_0$  &     0.000  &  0     &   0   &     0  &      & 52.6\footnotemark[1] \\ 
 2 &            & $^3$P$_1$  &  1691.806  & 1.502  &  1831 &  1.500 & 12   & 49.6 \\
 3 &            & $^3$P$_2$  &  3427.673  & 1.452  &  3557 &  1.447 & 16   & 50.1 \\
 4 & $5s^25p^2$ & $^1$D$_2$  &  8612.955  & 1.052  &  8860 &  1.053 & 1.0  & 54.2 \\
 5 & $5s^25p^2$ & $^1$S$_0$  & 17162.499  &   0    & 17496 &    0   & 0.15 & 70.5 \\
    	        	   	      	           	    	     	    
 6 & $5s^25p6s$ & $^3$P$^o_0$  & 34640.758  &  0     & 34380 & 0      &   & \\
 7 &            & $^3$P$^o_1$  & 34914.282  & 1.380  & 34622 &  1.371 &   & \\
 8 &            & $^3$P$^o_2$  & 38628.876  & 1.501  & 38392 &  1.513 &   & \\
 9 & $5s^25p6s$ & $^1$P$^o_1$  & 39257.053  & 1.121  & 39044 &  1.126 &   & \\
10 & $5s5p^3$   & $^5$S$^o_2$  & 39625.506  &        & 39997 &  1.978 &   & \\
\end{tabular}
\footnotetext[1]{Recommended value from Ref.~\cite{PolTab} is 53(6) $a_0^3$.}
\end{ruledtabular}
\end{table}

Finally, Table~\ref{t:SnPNC} presents the calculated $E1^{\rm PNC}$ and M1 amplitudes, as well as their ratios, for all transitions between states of the ground configuration. Notably, the value for the $^1$S$_0$–$^3$P$_1$ transition is only about three times smaller than that of the measured PNC transition in Pb.

This suggests that experimental measurements of the PNC effect in Sn should be similarly feasible.

\begin{table}
 \caption{\label{t:SnPNC}
 Reduced matrix elements   (RME)  of  $E1^{\rm PNC}$ and M1 amplitudes between states of the ground configuration of $^{120}$Sn. 
 $E1^{\rm PNC}$ (RME and $z$-component) are  presented  in units of $iea_B$. Parameter  $R={\rm Im}(E1^{\rm PNC})/M1$.}
\begin{ruledtabular}
\begin{tabular}{cccc cccc}
$N$ & 
\multicolumn{3}{c}{Transition}&
\multicolumn{2}{c}{$E1^{\rm PNC}$}&
\multicolumn{1}{c}{M1}&
\multicolumn{1}{c}{$R$}\\
&&&&\multicolumn{1}{c}{RME}&
\multicolumn{1}{c}{$z$}&
\multicolumn{1}{c}{[$\mu_B$]}& \\
\hline
1 & $^3$P$_0$ &-& $^3$P$_1$ & 1.72[-11] &  9.92[-12] &  1.387 &  3.40[-9]  \\
2 & $^1$S$_0$ &-& $^3$P$_1$ & 2.90[-11] &  1.67[-11] &  0.255 &  3.12[-8]  \\
3 & $^3$P$_2$ &-& $^3$P$_1$ & 1.81[-11] &  5.72[-12] &  1.493 &  3.32[-9]  \\
4 & $^1$D$_2$ &-& $^3$P$_1$ & 2.46[-11] &  7.79[-12] &  0.506 &  1.34[-8]  \\
5 & $^1$D$_2$ &-& $^3$P$_2$ & 7.97[-13] &  2.91[-13] &  0.830 &  2.63[-10] \\


\end{tabular}
\end{ruledtabular}
\end{table}

Finally, Table~\ref{t:mZ} shows the contribution of a $Z'$ boson to the PNC amplitudes in Sn as a function of $Z'$ mass. The units are chosen such that, in the limit $m_{Z'} \rightarrow \infty$, the $Z'$ contribution reduces to that of the Standard Model $Z$ boson. Therefore, the values in the first row of Table~\ref{t:mZ} are close to  the corresponding $E1^{\rm PNC}_z$ amplitudes presented in Table~\ref{t:SnPNC} (column marked as "$z$").

The use of these data for putting constraints on the strength of new interactions from the PNC measurements are discussed in detail in our previous papers~\cite{DFS17,DFV26}.

\begin{table}
 \caption{\label{t:mZ}
PNC amplitudes ($z$-components) induced by interaction (\ref{e:V12})
for various $Z'$ boson masses. The presented values for
the atomic PNC amplitudes are in terms of the parameter $-Q_W/N=(2\sqrt{2}Ag^A_e g^V_N)/(NG_Fm^2_{Z'})$ and in the units $iea_B$.
Numbers in square brackets stand for powers of ten.}

\begin{ruledtabular}
\begin{tabular}{l ccccc}
\multicolumn{1}{c}{$m_{Z'}$}& & & & & \\
\multicolumn{1}{c}{(eV)}&
\multicolumn{1}{c}{$^3$P$_0 - ^3$P$_1$}&
\multicolumn{1}{c}{$^1$S$_0 - ^3$P$_1$}&
\multicolumn{1}{c}{$^3$P$_2 - ^3$P$_1$}&
\multicolumn{1}{c}{$^1$D$_2 - ^3$P$_1$}&
\multicolumn{1}{c}{$^1$D$_2 - ^3$P$_2$}\\

\hline
$10^{10}$ & 1.02[-11] & 1.72[-11] & 5.87[-12] & 7.98[-12] & -1.15[-11] \\
$10^{9}$  & 1.02[-11] & 1.72[-11] & 5.87[-12] & 7.97[-12] & -1.15[-11] \\
$10^{8}$  & 1.00[-11] & 1.68[-11] & 5.75[-12] & 7.81[-12] & -1.13[-11] \\
$10^{7}$  & 7.90[-12] & 1.33[-11] & 4.54[-12] & 6.16[-12] & -8.90[-12] \\
$10^{6}$  & 3.94[-12] & 6.61[-12] & 2.26[-12] & 3.07[-12] & -4.44[-12] \\
$10^{5}$  & 3.33[-13] & 5.57[-13] & 1.90[-13] & 7.98[-12] & -3.75[-13] \\
$10^{4}$  & 4.23[-15] & 7.19[-15] & 2.45[-15] & 3.19[-15] & -4.83[-15] \\
$10^{3}$  & 3.34[-17] & 5.62[-17] & 1.92[-17] & 2.61[-17] & -3.76[-17] \\
$10^{2}$  & 3.13[-19] & 5.26[-19] & 1.80[-19] & 2.46[-19] & -3.52[-19] \\
$10^{1}$  & 3.12[-21] & 5.24[-21] & 1.79[-21] & 2.45[-21] & -3.51[-21] \\

\end{tabular}
\end{ruledtabular}
\end{table}

\subsection{Neutron skin}

As discussed above, the accuracy of the present calculations of the PNC effect in Sn is about 5\%. In this respect, Sn cannot compete with Cs, where the accuracy reaches about 0.5\%~\cite{DereviankoCsPNC,CsPNC12}. The main source of uncertainty is the incomplete treatment of correlations. Further improvement in accuracy is challenging and is unlikely to reach the level achieved for Cs in the foreseeable future.

However, tin has another important advantage: the large number of stable isotopes. Ratios of PNC effects between different isotopes are largely independent of atomic calculations and can provide valuable information about neutron and proton distributions in the nucleus, as well as place constraints on physics beyond the Standard Model.

\begin{table}
 \caption{\label{t:beta} 
 Parameters $k$ and $\beta$ (see Eq.~(\ref{e:r})) for the $\langle 5s_{1/2} | \hat W| 5p_{1/2} \rangle$ matrix elements in
 tin obtained by fitting numerical calculations.}
\begin{ruledtabular}
\begin{tabular}{ldd}
&\multicolumn{1}{c}{Fermi}&
\multicolumn{1}{c}{RMS}\\
\hline
$k$       & 0.0561 & 0.0672 \\ 
$\beta$ & -0.1224 &  -0.1482 \\
\end{tabular}
\end{ruledtabular}
\end{table}

\begin{table*}
\caption{\label{t:r} 
Parameters on nuclear proton and neutron distributions for some isotopes of Sn.
The last column presents radius of the proton Fermi distribution, which corresponds to the RMS data of Angeli and Marinova~\cite{Marinova} (see formula (\ref{e:rms})).
}
\begin{ruledtabular}
\begin{tabular}{cc cccccc}
\multicolumn{1}{c}{$A$}&
\multicolumn{1}{c}{$N$}&
\multicolumn{2}{c}{$\sqrt{\langle r^2 \rangle_p}$ \ fm}&
\multicolumn{1}{c}{$\sqrt{\langle r^2 \rangle_n}$ \ fm}&
\multicolumn{2}{c}{$\Delta r_{np}=\sqrt{\langle r^2 \rangle_n}-\sqrt{\langle r^2 \rangle_p}$ \ fm}&
\multicolumn{1}{c}{$R_p$ \ fm}\\
&&\multicolumn{1}{c}{Ref.~\cite{Marinova}}&
\multicolumn{1}{c}{Ref.~\cite{Terashima}}&
\multicolumn{1}{c}{Ref.~\cite{Terashima}}&
\multicolumn{1}{c}{Ref.~\cite{Terashima}}&
\multicolumn{1}{c}{Ref.~\cite{Tagami}}& \\
\hline
116 & 66 & 4.6250(19) & 4.562(3) & 4.672(18) & 0.110(18) & 0.118(21) & 5.4169(22)  \\
118 & 68 & 4.6393(19) & 4.575(3) & 4.720(16) & 0.145(16) & 0.112(21) & 5.4373(22) \\
120 & 70 & 4.6519(21) & 4.589(3) & 4.736(33) & 0.147(33) & 0.124(21) & 5.4552(25) \\
122 & 72 & 4.6634(22) & 4.602(3) & 4.748(16) & 0.146(16) & 0.122(24) & 5.4715(26) \\
124 & 74 & 4.6735(23) & 4.615(3) & 4.800(17) & 0.185(17) & 0.156(22) & 5.4859(27) \\
\end{tabular}
\end{ruledtabular}
\end{table*}

In the case when neutron and proton distributions in the nucleus are different, the 
nuclear density in (\ref{e:hpnc}) can be replaced by weighted sum of proton and neutron densities
\begin{equation}
\tilde \rho(r) = (q_n N  \rho_n(r) + q_p Z \rho_p(r))/Q_W,
\end{equation}
where $q_n = -0.98207$, $q_p = 0.071918$ are the neutron and proton weak charges, and the neutron  and proton densities  $\rho_n(r)$ and  $\rho_p(r)$ are normalised to 1.  
The radius of the Fermi distribution for neutrons is usually larger than that for protons. This difference is called neutron skin.

The PNC amplitude is proportional to the matrix elements between $s_{1/2}$ and $p_{1/2}$ states
\begin{equation}\label{e:me}
\mathcal{M} = \langle ns_{1/2} | \hat W| np_{1/2} \rangle.
\end{equation}
The ratio of the PNC amplitudes for a pair of isotopes is equal to the ratio of weak matrix elements, which  may be fitted by the following formula:
\begin{equation}\label{e:r}
 \frac{\mathcal{M}_1}{\mathcal{M}_2} = \frac{Q'_{W1}}{Q'_{W2}} 
\left[1 + k\left(\frac{\delta R_1}{R_1} - \frac{\delta R_2}{R_2}\right) \right] \left(\frac{R_1}{R_2}\right)^{\beta}.
\end{equation}
 This formula is motivated by analytical estimates of the dependence of the weak  matrix elements on the  proton and neutron distribution radii \cite{Brown,Viatkina}. { Here, 
 \begin{equation}\label{e:r2}
 Q_W'= (q_p + \Delta_p) Z + (q_n + \Delta_n) N=Q_W + \Delta Q_W, 
 \end{equation} 
 where $\Delta_p$ and $\Delta_n$ 
 denote hypothetical contributions to the proton and neutron  weak  charges arising from a new contact interaction beyond the Standard Model.}
Note, however, that there may be additional contributions to the PNC amplitudes that cannot be reduced to a simple correction to the weak nuclear charge.
For example, the exchange of a hypothetical new $Z'$ boson can generate such a contribution. In the case of a low $Z'$ mass, the interaction becomes long-ranged.

The parameter $R$ in (\ref{e:r}) is the radius of the proton distribution in the nucleus,
$\delta R = R_n - R_p$ is the difference between neutron and proton nuclear radii (neutron skin).
Parameters $k$ and $\beta$ can be found from fitting numerical calculations. 
{
The changes in the ratio Eq. (\ref{e:r}) due to the parameters $\beta$ and $k$ are small and practically independent. Therefore, the two parameters can be determined separately.

To determine $\beta$, we disregard both the neutron-skin effect and the variation of the weak nuclear charge $Q_W$, so that 
\[
\frac{\mathcal{M}_1}{\mathcal{M}_2}=\left(\frac{R_1}{R_2}\right)^{\beta},
\]
which gives
\[
\beta=\frac{\ln(\mathcal{M}_1/\mathcal{M}_2)}{\ln(R_1/R_2)}.
\]
We determine $\beta$ by varying the rms radius of the proton distribution in the atomic-structure calculations and evaluating the corresponding weak-interaction matrix elements. Several values of the nuclear radius are used to verify that the dependence of the matrix elements on $\ln R$ is linear, as assumed in the above expression.

Similarly, to determine $k$, we use the relation
\[
\frac{\delta \mathcal{M}}{\mathcal{M}}=k\frac{\delta R}{R},
\]
where $\delta R$ is the change in the rms radius of the neutron distribution. In this case, the neutron radius is varied while all other parameters are kept fixed. Again, several values of $\delta R$ are considered to verify the linear dependence assumed in the definition of $k$.


}

Note that $R$ in (\ref{e:r}) can be either the radius of the Fermi distribution ($R_p$) or the root mean square radius ($R_{\rm RMS}$). These two nuclear parameters are related to each other to very high accuracy by \cite{Viatkina}
\begin{equation}\label{e:rms}
R_{\rm RMS}^2 = \frac{3}{5}R_p^2 + 3.785 \ {\rm fm^2}.
\end{equation}
Note that if $R_p = 1.1 A^{1/3}$~fm, then Eq.~(\ref{e:rms}) yields values that agree with the compilation by Angeli and Marinova~\cite{Marinova} to better than 1\% accuracy.

The values of $k$ and $\beta$ depend on whether $R_p$ or $R_{\rm RMS}$ are used in (\ref{e:r}). 
The numbers for tin ($Z=50$) obtained by fitting numerical calculations are presented in Table~\ref{t:beta}.
Analytical consideration for the case of the uniformly charged nuclear ball \cite{Brown,Viatkina}, these relations  give $\beta = 2\gamma-2$, where $\gamma=\sqrt{1-(Z\alpha)^2}$.
For the case of Sn ($Z=50$), $\gamma=0.931$ and $\beta= 2\gamma-2= -0.138$. Numerical fitting gives close but presumably more accurate values of $\beta$ (see Table~\ref{t:beta}).

{
Equation~(\ref{e:r}), together with the data from Table~\ref{t:beta}, can be used to analyze the results of PNC measurements in terms of the neutron skin or new physics. Dividing measured values  of the ratio of $\mathcal{M}_1/\mathcal{M}_2$ by two last factors depending on the neutron skins and charge distribution radii, we obtain the ratio of the effective weak charges $Q'_{W1}/Q'_{W2}$. The effective weak charge is a linear function of the neutron number $N$, see Eq.~(\ref{e:r2}). After fitting the results of the measurements of PNC effect ratios for all available isotopes  with a linear function $Q'_W=a +b N$ and subtracting the standard model weak charge values, we obtain the new physics contribution to the weak charge. This procedure has been realised in Refs.~\cite{Antypas2018,Antypas2019}, where   measurements of the PNC effects for Yb isotopes were performed. The effect of  the neutron skin was neglected in that work, since the experimental errors of $\sim 1$\% in the measured PNC effects  exceeded the neutron skin effect.   

To perform more sensitieve measurements of new physics, one  should include this effect. In this case both experimental errors in the PNC effects and errors in the neutron skin values propagate to the error in the measured value of any new physics effect.   For convenience of further references, we extract from (\ref{e:r}) the neutron skin factor $F_{ns}$,
\begin{equation}\label{e:fns}
F_{ns} = k\left(\frac{\delta R_1}{R_1} - \frac{\delta R_2}{R_2}\right),
\end{equation}
For an efficient search for new physics, the $F_{ns}$ factor should be either small or known to sufficient accuracy.
Fortunately, the values of the neutron skin are known for several Sn isotopes; these data are summarized in Table~\ref{t:r}.
Table \ref{t:ns} presents the values of $F_{ns}$ calculated using the data from Table~\ref{t:r}.

The errors were evaluated under the assumption that the uncertainties in the neutron-skin values for different isotopes are independent. However, nuclear calculations~\cite{Brown} show that these uncertainties are in fact significantly correlated and this  may lead to a partial cancellation of errors in the difference of the neutron skins. This correlation may  reduce the uncertainty in $F_{ns}$ by a factor of 4 to 10. The reason is that the core neutron distribution is largely insensitive to the addition of extra neutrons to the nucleus. Consequently, the uncertainty in the core neutron distribution cancels to a large extent when one considers the difference between the neutron distribution radii of isotopes. 

The uncertainty associated with $F_{ns}$ is of the order of $10^{-4}$ (see Table~\ref{t:ns}). One  can improve accuracy in the estimate of the contribution of new physics using measurements of PNC effects on several isotopes and fitting contribution of new physics for all of them. Here the large number of Sn stable isotopes gives an important advantage. Another advantage is that the neutron skin effects, proportional to $Z^2 \alpha^2$,  are smaller in Sn compared to in heavier atoms Cs, Dy, Sm, Yb, Tl, Pb, Bi, where previous measurements of PNC effects were performed. Therefore, the neutron skin "floor" in the accuracy of  the new physics measurement may, in principle,  be several times smaller than  $10^{-4}$, using existing neutron skin data. New nuclear calculations of the differences  of the neutron skin thickness  for different Sn isotopes, combined with the neutron skin  measurements, may give significant further improvements.}

\begin{table}
\caption{\label{t:ns} 
Neutron skin factor $F_{ns}$ (\ref{e:fns}) in units $10^{-4}$ for different sets of data from Table~\ref{t:r}. }
\begin{ruledtabular}
\begin{tabular}{cc rrr}
\multicolumn{2}{c}{Pair of}&
\multicolumn{1}{c}{\cite{Marinova,Tagami}}&
\multicolumn{1}{c}{\cite{Marinova,Terashima}}&
\multicolumn{1}{c}{\cite{Terashima}}\\
\multicolumn{2}{c}{issotopes}&&& \\
\hline
  118 & 116 & -0.9  $\pm$ 4.3 &  5.0  $\pm$ 3.5 &  5.1  $\pm$ 3.5 \\
  120 & 116 &  0.8  $\pm$ 4.3 &  5.3  $\pm$ 5.4 &  5.3  $\pm$ 5.5 \\
  122 & 116 &  0.4  $\pm$ 4.6 &  5.1  $\pm$ 3.5 &  5.1  $\pm$ 3.5 \\
  124 & 116 &  5.3  $\pm$ 4.4 & 10.6  $\pm$ 3.6 & 10.7  $\pm$ 3.6 \\
  120 & 118 &  1.7  $\pm$ 4.3 &  0.2  $\pm$ 5.3 &  0.2  $\pm$ 5.4 \\
  122 & 118 &  1.4  $\pm$ 4.6 &  0.0  $\pm$ 3.3 &  0.0  $\pm$ 3.3 \\
  124 & 118 &  6.2  $\pm$ 4.4 &  5.6  $\pm$ 3.4 &  5.6  $\pm$ 3.4 \\
  122 & 120 & -0.3  $\pm$ 4.6 & -0.2  $\pm$ 5.3 & -0.2  $\pm$ 5.4 \\
  124 & 120 &  4.5  $\pm$ 4.4 &  5.4  $\pm$ 5.4 &  5.4  $\pm$ 5.4 \\
  124 & 122 &  4.9  $\pm$ 4.7 &  5.6  $\pm$ 3.4 &  5.6  $\pm$ 3.4 \\
\end{tabular}
\end{ruledtabular}
\end{table}

\subsection{Experimental Scheme to Measure PNC in Sn}
In this section, we briefly describe an experimental scheme for measuring PNC on the ${^3}\!P_1 \leftrightarrow ^{1}\!S_0$ transition in atomic Sn, with natural linewidth $\Gamma \approx 2\pi\times 1.1$~Hz (corresponding to lifetime $\tau_{^{1\!}S_0} \approx 0.15$~s) 
We go on to discuss the potential precision of such a measurement.

{Based on extensive simulations of laser cooling and trapping of Sn atoms \cite{SnSimsPaper}, we anticipate achieving phase-space density (temperature) sufficiently high (low) to load over $\!10^4$ atoms into a 1-D optical lattice, formed in a high-finesse optical cavity. 
Trapping in a deep lattice will allow for Raman sideband cooling and/or EIT cooling into the ground state of axial motion in the lattice.
Light near resonance with the PNC transition itself, 
at wavelength $\lambda_1 = 646$ nm, can form the lattice for both states in the transition. Alternatively, primary trapping can be performed with light at $\lambda_{1/3}=3\lambda_1$, which is commensurate with the $\lambda_1$ lattice, in the same cavity. Here we defined $\lambda_f = \lambda_1/f$.

The electric field antinodes of the PNC transition standing wave coincide with the trapped atom location. However, the magnetic field of the light has zeros at the antinodes. This strongly suppresses the M1 transition amplitude that would otherwise dominate the measurement signal.

In this configuration, a measurement scheme analogous to that proposed in Ref.~\cite{Fortson-Ion} can be employed.  Here, the actual measurement is of an AC Stark shift, differential between the Zeeman sublevels $m_J = \pm 1$ of the $^3P_1$ state, caused by interference between an allowed amplitude and the PNC amplitude, each driving the $^3P_1 
\leftrightarrow~\!^1S_0$ transition.
In more detail: a laser is tuned to near resonance with the transition (with detuning $\delta$), and the allowed amplitude is driven with Rabi frequency $\Omega_0 \gg \delta$. Simultaneously, the PNC amplitude is driven with Rabi frequency $\Omega^{\rm {PNC}}_{m_J}$. Here, $\Omega^{\rm {PNC}}_{m_J} = m_J E1^{\rm PNC} E_0/2$, where $E_0$ is the peak magnitude of the electric field of the near-resonant light in the cavity. Interference between the two drive amplitudes leads to an AC Stark shift, $\delta\omega_{m_J}$, of the atomic levels, given by $\delta\omega_{m_J} = \delta/2 + \sqrt{(\Omega_0 + \Omega^{\rm {PNC}}_{m_J})^2} \approx \delta/2 + \Omega_0+ \Omega^{\rm {PNC}}_{m_J}$. Finally, this leads to a differential shift between the Zeeman sublevels, $\Delta(\delta\omega) = \delta\omega_{1}-\delta\omega_{-1} = 2\Omega^{\rm {PNC}}_{m_J}$, which can be measured with standard methods of Ramsey interferometry.

The parity-allowed amplitude here can be applied by using a two-photon transition 
\cite{Elliot-PNCScheme}. Unlike in Ref. \cite{Elliot-PNCScheme}, the light driving the two-photon amplitude cannot be at wavelength $\lambda_{1/2}$, because a two-photon transitions between states with $J=0$ and $J'=1$ is forbidden for photons of the same frequency \cite{demille1999search}.  Instead, we plan to use different combinations such as $\lambda_{1/3} + \lambda_{2/3}$ or $\lambda_{4/3} - \lambda_{1/3}$.  The two-photon amplitudes are estimated to be sufficiently weak that light of high intensity---comparable or greater to that at $\lambda_{1}$---will be needed.  Because this can affect the trapping conditions, we calculated the polarizabilities of the two states of interest at all these frequencies  (see Table \ref{t:pol}).

\begin{table}
 \caption{\label{t:pol}
 Dynamic polarizabilities of the $^3$P$_1$ and $^1$S$_0$ states of Sn
 calculated at fractions of the transition frequency $\omega_0$ = 15472~cm$^{-1}$ = 0.0705~a.u.}
\begin{ruledtabular}
\begin{tabular}{r c rrrr}
\multicolumn{2}{c}{$\omega$}&
\multicolumn{3}{c}{$^3$P$_1$}&
\multicolumn{1}{c}{$^1$S$_0$}\\

\multicolumn{1}{c}{$k\omega_0$}&
\multicolumn{1}{c}{[a.u.]}&
\multicolumn{1}{c}{scalar}&
\multicolumn{1}{c}{vector}&
\multicolumn{1}{c}{tensor}&
\multicolumn{1}{c}{scalar}\\
\hline
                      & 0.00000 & 52.6  &   0.00 & -2.82 &  70.5 \\
$\frac{1}{3}\omega_0$ &	0.02350 & 53.2  &  -0.47 & -2.90 &  71.1 \\
$\frac{1}{2}\omega_0$ &	0.03525 & 54.1  &  -0.82 & -3.01 &  74.3 \\ 
$\frac{2}{3}\omega_0$ &	0.04700 & 55.3  &  -1.23 & -3.19 &  77.9 \\
$           \omega_0$ &	0.07050 & 59.4  &  -2.89 & -3.78 &  99.3 \\
$\frac{4}{3}\omega_0$ &	0.09400 & 67.0  &  -7.65 & -5.03 & 139.1 \\
\end{tabular}
\end{ruledtabular}
\end{table}

The maximum value of $E_0$, and hence $|\Omega^{\rm PNC}|$, is set by demanding that the loss rate due to scattering, $\Gamma_{\rm sc}$, be smaller than the spontaneous emission rate. From our calculations of polarizabilities and scattering rates, we find that light intensity $I_{\rm 646} \approx 1\times 10^7$~W/cm$^2$ (corresponding to $E_0 \approx 1\times10^5$~V/cm) yields $\Gamma_{\rm sc} \approx 0.2 \Gamma$, trap depth $U \approx k_B\times 1.5$~mK, and Rabi frequency $|\Omega^{\rm PNC}| \approx 2\pi\times 1.5$~Hz. Using a resonant optical cavity with finesse $\mathcal{F} = 1000$, this intensity is easily achieved.  More details about the measurement scheme will be provided in an upcoming publication.

This approach can yield a statistical fractional uncertainty on $E1^{\rm PNC}$, given by \cite{Fortson-Ion}:
\begin{equation}
    \frac{\delta (E1^{\rm PNC})}{E1^{\rm PNC}} \cong \frac{1}{\Omega^{\rm PNC}}\frac{1}{f\sqrt{N\tau T}}. \nonumber
\end{equation}
Here, $T$ is the total measurement time; $\tau \approx 2/\Gamma \approx 0.2$~s is the lifetime of the state dressed by the strong parity-allowed drive (including the effects of scattering); $N\approx 10^4$ is the number of atoms in the trap; and $f$ is a factor that accounts for various experimental inefficiencies. With a reasonable estimate of $f=0.3$, we expect a relative statistical uncertainty of $\delta (E1^{\rm PNC})/{E1^{\rm PNC}} \approx 1\times10^{-5}$ for a single isotope, with $T \approx 100$~hours.

We next consder how this uncertainty is related to a new physics-induced deviation from the SM value. Here, we assume the same couplings to the $Z'$ boson as earlier (see Eqs.~(\ref{e:Lint}-\ref{e:V12})). In the limit of a very massive $Z'$ boson, the Yukawa-like interaction becomes a contact interaction, and the contribution to the weak nuclear charge from the $Z'$ boson is given by Eq.~(\ref{e:r2}). The fractional shift in the isotopic ratio of PNC amplitudes $\mathcal{R} \equiv \mathcal{M}_1/\mathcal{M}_2$ is then:
\begin{equation}
    \delta \mathcal{R} \approx \frac{\Delta Q_{W1}}{Q_{W1}} - \frac{\Delta Q_{W2}}{Q_{W2}}, \nonumber
\end{equation}
where the subscripts 1 and 2 indicate the two isotopes being compared.

We relate $\Delta Q_W$ to the axial electron coupling, vector nucleon coupling, and hypothetical $Z'$ boson mass via:
\begin{equation}
    -\frac{\Delta Q_W}{N} = \frac{2\sqrt{2}Ag_e^A g_N^V}{N G_F m_{Z'}^2}. \nonumber
\end{equation}

Choosing isotope 1 = $^{116}$Sn and isotope 2 = $^{124}$Sn, assuming an isoscalar coupling, and using Eq.~(\ref{e:qwn}) as the corrected expression for weak charge, we find:
\begin{equation}
    \delta \mathcal{R} \approx 2.4
    \times 10^{-4} \left(\frac{g_e^A g_N^V/m_{Z'}^2}{10^{-8}\,\rm{GeV}^{-2}} \right). \nonumber
\end{equation}


The current bound on $|g_e^Ag_N^V|/m_{Z'}^2$ from atomic PNC experiments, assuming a heavy $Z'$ boson, is given by the measurement in $^{133}$Cs ~\cite{DFS17}: 
\begin{equation}
|g_e^Ag_N^V|/m_{Z'}^2 < 3.9 \times 10^{-8}~\mathrm{GeV}^{-2}.
\nonumber
\end{equation} 
From this bound, we conclude that
    $\delta \mathcal{R} < 9.4 \times 10^{-4}$.
This is well above our projected experimental uncertainty of $\delta\mathcal{R}\sim \sqrt{2} \times 10^{-5}$. Moreover,  combination of the Cs single-isotope PNC measurement  with future Sn PNC measurement on many isotopes will allow one to measure  proton and neutron interaction constants  separately.  Also, for a Standard Model-like $Z'$, with $g_e^A = 1/2$ and $g_N^V \approx 1/4$, the proposed uncertainty would yield sensitivity to $m_{Z'} \lesssim 14$~TeV, well above direct bounds from the LHC \cite{SM2026}.

\section{Conclusion}

We have calculated the parity non-conservation amplitudes, M1 amplitudes, and their ratios for all transitions between states of the ground configuration of tin. The calculations were performed using a CI+coupled clusters+CIPT method and benchmarked against the well-studied PNC transition in Pb. The good agreement obtained for Pb gives confidence that the predicted PNC amplitudes for Sn are reliable at the level needed for experimental guidance.

Among the transitions considered, the $^1$S$_0$-{}$^3$P$_1$ transition has the largest PNC effect and is the most promising for observation. Its magnitude is only a few times smaller than that of the measured PNC transition in Pb, which suggests that an experimental study in Sn should be feasible. Moreover, tin satisfies all criteria laid out by Fortson \cite{Fortson-Ion} for a highly precise measurement enabled by measuring $E1^{\rm PNC}$ on an ultra-narrow transition.

Another key advantage of tin is the large number of stable isotopes, which makes it possible to measure ratios of PNC amplitudes for different isotopes. In such ratios, the electronic-structure factor cancels to a large extent, substantially reducing the dependence on atomic calculations and enhancing sensitivity to nuclear effects and possible new physics. We have examined the influence of the neutron skin on these ratios using available data for Sn isotopes and found that  the uncertainty associated with the neutron skin can be reduced to the 
$\sim 10^{-4}$ level relative to the isotopic variation of the PNC effect. 

We have calculated the contribution of an additional $Z'$ boson to PNC effects. The ratio of the low mass $Z'$ effect to the standard model $Z$ boson effect in Sn is bigger than this ratio in heavier atoms.

We therefore conclude that isotope-ratio measurements of PNC in tin provide a realistic and theoretically clean pathway to precision tests of the Standard Model and to searches for new parity-violating interactions beyond it.

\begin{acknowledgments}

This work was supported by the Australian Research Council Grant No. DP230101058.
\end{acknowledgments}


\begin{thebibliography}{42}
\expandafter\ifx\csname natexlab\endcsname\relax\def\natexlab#1{#1}\fi
\expandafter\ifx\csname bibnamefont\endcsname\relax
  \def\bibnamefont#1{#1}\fi
\expandafter\ifx\csname bibfnamefont\endcsname\relax
  \def\bibfnamefont#1{#1}\fi
\expandafter\ifx\csname citenamefont\endcsname\relax
  \def\citenamefont#1{#1}\fi
\expandafter\ifx\csname url\endcsname\relax
  \def\url#1{\texttt{#1}}\fi
\expandafter\ifx\csname urlprefix\endcsname\relax\def\urlprefix{URL }\fi
\providecommand{\bibinfo}[2]{#2}
\providecommand{\eprint}[2][]{\url{#2}}

\bibitem[{\citenamefont{Wood et~al.}(1997)\citenamefont{Wood, Bennett, Cho,
  Masterson, Roberts, Tanner, and Wieman}}]{Wood}
\bibinfo{author}{\bibfnamefont{C.~S.} \bibnamefont{Wood}},
  \bibinfo{author}{\bibfnamefont{S.~C.} \bibnamefont{Bennett}},
  \bibinfo{author}{\bibfnamefont{D.}~\bibnamefont{Cho}},
  \bibinfo{author}{\bibfnamefont{B.~P.} \bibnamefont{Masterson}},
  \bibinfo{author}{\bibfnamefont{J.~L.} \bibnamefont{Roberts}},
  \bibinfo{author}{\bibfnamefont{C.~E.} \bibnamefont{Tanner}},
  \bibnamefont{and} \bibinfo{author}{\bibfnamefont{C.~E.}
  \bibnamefont{Wieman}}, \bibinfo{journal}{Science}
  \textbf{\bibinfo{volume}{275}}, \bibinfo{pages}{1759} (\bibinfo{year}{1997}).

\bibitem[{\citenamefont{Dzuba et~al.}(1989)\citenamefont{Dzuba, Flambaum, and
  Sushkov}}]{DzuFlaSus89a}
\bibinfo{author}{\bibfnamefont{V.~A.} \bibnamefont{Dzuba}},
  \bibinfo{author}{\bibfnamefont{V.~V.} \bibnamefont{Flambaum}},
  \bibnamefont{and} \bibinfo{author}{\bibfnamefont{O.~P.}
  \bibnamefont{Sushkov}}, \bibinfo{journal}{Phys. Lett. A}
  \textbf{\bibinfo{volume}{141}}, \bibinfo{pages}{147} (\bibinfo{year}{1989}).

\bibitem[{\citenamefont{Blundell et~al.}(1990)\citenamefont{Blundell, Johnson,
  and Sapirstein}}]{BluJohSap1990}
\bibinfo{author}{\bibfnamefont{S.~A.} \bibnamefont{Blundell}},
  \bibinfo{author}{\bibfnamefont{W.~R.} \bibnamefont{Johnson}},
  \bibnamefont{and}
  \bibinfo{author}{\bibfnamefont{J.}~\bibnamefont{Sapirstein}},
  \bibinfo{journal}{Phys. Rev. Lett.} \textbf{\bibinfo{volume}{65}},
  \bibinfo{pages}{1411} (\bibinfo{year}{1990}),
  \urlprefix\url{https://link.aps.org/doi/10.1103/PhysRevLett.65.1411}.

\bibitem[{\citenamefont{Dzuba et~al.}(2002)\citenamefont{Dzuba, Flambaum, and
  Ginges}}]{DzuFlaGin02}
\bibinfo{author}{\bibfnamefont{V.~A.} \bibnamefont{Dzuba}},
  \bibinfo{author}{\bibfnamefont{V.~V.} \bibnamefont{Flambaum}},
  \bibnamefont{and} \bibinfo{author}{\bibfnamefont{J.~S.~M.}
  \bibnamefont{Ginges}}, \bibinfo{journal}{Phys. Rev. D}
  \textbf{\bibinfo{volume}{66}}, \bibinfo{pages}{076013}
  (\bibinfo{year}{2002}).

\bibitem[{\citenamefont{Dzuba et~al.}(2012)\citenamefont{Dzuba, Berengut,
  Flambaum, and Roberts}}]{CsPNC12}
\bibinfo{author}{\bibfnamefont{V.~A.} \bibnamefont{Dzuba}},
  \bibinfo{author}{\bibfnamefont{J.~C.} \bibnamefont{Berengut}},
  \bibinfo{author}{\bibfnamefont{V.~V.} \bibnamefont{Flambaum}},
  \bibnamefont{and} \bibinfo{author}{\bibfnamefont{B.}~\bibnamefont{Roberts}},
  \bibinfo{journal}{Phys. Rev. Lett.} \textbf{\bibinfo{volume}{109}},
  \bibinfo{pages}{203003} (\bibinfo{year}{2012}).

\bibitem[{\citenamefont{Porsev et~al.}(2009)\citenamefont{Porsev, Beloy, and
  Derevianko}}]{DereviankoCsPNC}
\bibinfo{author}{\bibfnamefont{S.~G.} \bibnamefont{Porsev}},
  \bibinfo{author}{\bibfnamefont{K.}~\bibnamefont{Beloy}}, \bibnamefont{and}
  \bibinfo{author}{\bibfnamefont{A.}~\bibnamefont{Derevianko}},
  \bibinfo{journal}{Phys. Rev. Lett.} \textbf{\bibinfo{volume}{102}},
  \bibinfo{pages}{181601} (\bibinfo{year}{2009}).

\bibitem[{\citenamefont{Tanabashi et~al.}(2018)\citenamefont{Tanabashi,
  Hagiwara, Hikasa, and {\em et al}}}]{SM}
\bibinfo{author}{\bibfnamefont{M.}~\bibnamefont{Tanabashi}},
  \bibinfo{author}{\bibfnamefont{K.}~\bibnamefont{Hagiwara}},
  \bibinfo{author}{\bibfnamefont{K.}~\bibnamefont{Hikasa}}, \bibnamefont{and}
  \bibinfo{author}{\bibnamefont{{\em et al}}} (\bibinfo{collaboration}{Particle
  Data Group}), \bibinfo{journal}{Phys. Rev. D} \textbf{\bibinfo{volume}{98}},
  \bibinfo{pages}{030001} (\bibinfo{year}{2018}),
  \urlprefix\url{https://link.aps.org/doi/10.1103/PhysRevD.98.030001}.

\bibitem[{\citenamefont{Flambaum and Samsonov}(2026)}]{Samsonov}
\bibinfo{author}{\bibfnamefont{V.~V.} \bibnamefont{Flambaum}} \bibnamefont{and}
  \bibinfo{author}{\bibfnamefont{I.~B.} \bibnamefont{Samsonov}},
  \bibinfo{journal}{Phys. Rev. D} \textbf{\bibinfo{volume}{114}},
  \bibinfo{pages}{L011302} (\bibinfo{year}{2026}), \eprint{2602.22466},
  \urlprefix\url{https://link.aps.org/doi/10.1103/cb8l-bt9n}.

\bibitem[{\citenamefont{Dzuba et~al.}(1986)\citenamefont{Dzuba, Flambaum, and
  Khriplovich}}]{DzuFlaKhr86}
\bibinfo{author}{\bibfnamefont{V.~A.} \bibnamefont{Dzuba}},
  \bibinfo{author}{\bibfnamefont{V.~V.} \bibnamefont{Flambaum}},
  \bibnamefont{and} \bibinfo{author}{\bibfnamefont{I.~B.}
  \bibnamefont{Khriplovich}}, \bibinfo{journal}{Z. Phys. D}
  \textbf{\bibinfo{volume}{1}}, \bibinfo{pages}{243} (\bibinfo{year}{1986}).

\bibitem[{\citenamefont{Brown et~al.}(2009)\citenamefont{Brown, Derevianko, and
  Flambaum}}]{Brown}
\bibinfo{author}{\bibfnamefont{B.~A.} \bibnamefont{Brown}},
  \bibinfo{author}{\bibfnamefont{A.}~\bibnamefont{Derevianko}},
  \bibnamefont{and} \bibinfo{author}{\bibfnamefont{V.~V.}
  \bibnamefont{Flambaum}}, \bibinfo{journal}{Phys. Rev. C}
  \textbf{\bibinfo{volume}{79}}, \bibinfo{pages}{035501}
  (\bibinfo{year}{2009}).

\bibitem[{\citenamefont{Viatkina et~al.}(2019)\citenamefont{Viatkina, Antypas,
  Kozlov, Budker, and Flambaum}}]{Viatkina}
\bibinfo{author}{\bibfnamefont{A.~V.} \bibnamefont{Viatkina}},
  \bibinfo{author}{\bibfnamefont{D.}~\bibnamefont{Antypas}},
  \bibinfo{author}{\bibfnamefont{M.~G.} \bibnamefont{Kozlov}},
  \bibinfo{author}{\bibfnamefont{D.}~\bibnamefont{Budker}}, \bibnamefont{and}
  \bibinfo{author}{\bibfnamefont{V.~V.} \bibnamefont{Flambaum}},
  \bibinfo{journal}{Phys. Rev. C} \textbf{\bibinfo{volume}{100}},
  \bibinfo{pages}{034318} (\bibinfo{year}{2019}).

\bibitem[{\citenamefont{Antypas et~al.}(2018)\citenamefont{Antypas, Fabricant,
  Stalnaker, Tsigutkin, Flambaum, and Budker}}]{Antypas2018}
\bibinfo{author}{\bibfnamefont{D.}~\bibnamefont{Antypas}},
  \bibinfo{author}{\bibfnamefont{A.~M.} \bibnamefont{Fabricant}},
  \bibinfo{author}{\bibfnamefont{J.~E.} \bibnamefont{Stalnaker}},
  \bibinfo{author}{\bibfnamefont{K.}~\bibnamefont{Tsigutkin}},
  \bibinfo{author}{\bibfnamefont{V.~V.} \bibnamefont{Flambaum}},
  \bibnamefont{and} \bibinfo{author}{\bibfnamefont{D.}~\bibnamefont{Budker}},
  \bibinfo{journal}{Nature Physics} \textbf{\bibinfo{volume}{15}},
  \bibinfo{pages}{120} (\bibinfo{year}{2018}),
  \urlprefix\url{https://doi.org/10.1038/s41567-018-0312-8}.

\bibitem[{\citenamefont{Antypas et~al.}(2019)\citenamefont{Antypas, Fabricant,
  Stalnaker, Tsigutkin, Flambaum, and Budker}}]{Antypas2019}
\bibinfo{author}{\bibfnamefont{D.}~\bibnamefont{Antypas}},
  \bibinfo{author}{\bibfnamefont{A.~M.} \bibnamefont{Fabricant}},
  \bibinfo{author}{\bibfnamefont{J.~E.} \bibnamefont{Stalnaker}},
  \bibinfo{author}{\bibfnamefont{K.}~\bibnamefont{Tsigutkin}},
  \bibinfo{author}{\bibfnamefont{V.~V.} \bibnamefont{Flambaum}},
  \bibnamefont{and} \bibinfo{author}{\bibfnamefont{D.}~\bibnamefont{Budker}},
  \bibinfo{journal}{Phys. Rev. A} \textbf{\bibinfo{volume}{100}},
  \bibinfo{pages}{012503} (\bibinfo{year}{2019}),
  \urlprefix\url{https://link.aps.org/doi/10.1103/PhysRevA.100.012503}.

\bibitem[{\citenamefont{Zhang et~al.}(2016)\citenamefont{Zhang, Collister,
  Shiells, Tandecki, Aubin, Behr, Gomez, Gorelov, Gwinner, Orozco
  et~al.}}]{Orozco}
\bibinfo{author}{\bibfnamefont{J.}~\bibnamefont{Zhang}},
  \bibinfo{author}{\bibfnamefont{R.}~\bibnamefont{Collister}},
  \bibinfo{author}{\bibfnamefont{K.}~\bibnamefont{Shiells}},
  \bibinfo{author}{\bibfnamefont{M.}~\bibnamefont{Tandecki}},
  \bibinfo{author}{\bibfnamefont{S.}~\bibnamefont{Aubin}},
  \bibinfo{author}{\bibfnamefont{J.~A.} \bibnamefont{Behr}},
  \bibinfo{author}{\bibfnamefont{E.}~\bibnamefont{Gomez}},
  \bibinfo{author}{\bibfnamefont{A.}~\bibnamefont{Gorelov}},
  \bibinfo{author}{\bibfnamefont{G.}~\bibnamefont{Gwinner}},
  \bibinfo{author}{\bibfnamefont{L.~A.} \bibnamefont{Orozco}},
  \bibnamefont{et~al.}, \bibinfo{journal}{Hyperfine Interaction}
  \textbf{\bibinfo{volume}{237}}, \bibinfo{pages}{150} (\bibinfo{year}{2016}).

\bibitem[{\citenamefont{Fortson et~al.}(1990)\citenamefont{Fortson, Pang, and
  Wilets}}]{Fortson}
\bibinfo{author}{\bibfnamefont{E.~N.} \bibnamefont{Fortson}},
  \bibinfo{author}{\bibfnamefont{Y.}~\bibnamefont{Pang}}, \bibnamefont{and}
  \bibinfo{author}{\bibfnamefont{L.}~\bibnamefont{Wilets}},
  \bibinfo{journal}{Phys. Rev. Lett.} \textbf{\bibinfo{volume}{65}},
  \bibinfo{pages}{2857} (\bibinfo{year}{1990}).

\bibitem[{\citenamefont{Trzci\'{n}ska et~al.}(2001)\citenamefont{Trzci\'{n}ska,
  Jastrz\c{e}bski, Lubi\'{n}ski, Hartmann, Schmidt, von Egidy, and
  Klos}}]{Trzcinska}
\bibinfo{author}{\bibfnamefont{A.}~\bibnamefont{Trzci\'{n}ska}},
  \bibinfo{author}{\bibfnamefont{J.}~\bibnamefont{Jastrz\c{e}bski}},
  \bibinfo{author}{\bibfnamefont{P.}~\bibnamefont{Lubi\'{n}ski}},
  \bibinfo{author}{\bibfnamefont{F.~J.} \bibnamefont{Hartmann}},
  \bibinfo{author}{\bibfnamefont{R.}~\bibnamefont{Schmidt}},
  \bibinfo{author}{\bibfnamefont{T.}~\bibnamefont{von Egidy}},
  \bibnamefont{and} \bibinfo{author}{\bibfnamefont{B.}~\bibnamefont{Klos}},
  \bibinfo{journal}{Phys. Rev. Lett.} \textbf{\bibinfo{volume}{87}},
  \bibinfo{pages}{08251} (\bibinfo{year}{2001}).

\bibitem[{\citenamefont{Terashima et~al.}(2008)\citenamefont{Terashima,
  Sakaguchi, Takeda, Ishikawa, Itoh, Kawabata, Murakami, Uchida, Yasuda, and
  et~al.}}]{Terashima}
\bibinfo{author}{\bibfnamefont{S.}~\bibnamefont{Terashima}},
  \bibinfo{author}{\bibfnamefont{H.}~\bibnamefont{Sakaguchi}},
  \bibinfo{author}{\bibfnamefont{H.}~\bibnamefont{Takeda}},
  \bibinfo{author}{\bibfnamefont{T.}~\bibnamefont{Ishikawa}},
  \bibinfo{author}{\bibfnamefont{M.}~\bibnamefont{Itoh}},
  \bibinfo{author}{\bibfnamefont{T.}~\bibnamefont{Kawabata}},
  \bibinfo{author}{\bibfnamefont{T.}~\bibnamefont{Murakami}},
  \bibinfo{author}{\bibfnamefont{M.}~\bibnamefont{Uchida}},
  \bibinfo{author}{\bibfnamefont{Y.}~\bibnamefont{Yasuda}}, \bibnamefont{and}
  \bibinfo{author}{\bibnamefont{et~al.}}, \bibinfo{journal}{Phys. Rev. C}
  \textbf{\bibinfo{volume}{77}}, \bibinfo{pages}{024317}
  (\bibinfo{year}{2008}).

\bibitem[{\citenamefont{Roca-Maza et~al.}(2008)\citenamefont{Roca-Maza, Vinas,
  Centelles, Agrawal, Col\'{o}, Paar, Piekarewicz, and Vretenar}}]{Roca-Maza}
\bibinfo{author}{\bibfnamefont{X.}~\bibnamefont{Roca-Maza}},
  \bibinfo{author}{\bibfnamefont{X.}~\bibnamefont{Vinas}},
  \bibinfo{author}{\bibfnamefont{M.}~\bibnamefont{Centelles}},
  \bibinfo{author}{\bibfnamefont{B.~K.} \bibnamefont{Agrawal}},
  \bibinfo{author}{\bibfnamefont{G.}~\bibnamefont{Col\'{o}}},
  \bibinfo{author}{\bibfnamefont{N.}~\bibnamefont{Paar}},
  \bibinfo{author}{\bibfnamefont{J.}~\bibnamefont{Piekarewicz}},
  \bibnamefont{and} \bibinfo{author}{\bibfnamefont{D.}~\bibnamefont{Vretenar}},
  \bibinfo{journal}{Phys. Rev. C} \textbf{\bibinfo{volume}{77}},
  \bibinfo{pages}{064304} (\bibinfo{year}{2008}).

\bibitem[{\citenamefont{Tagami et~al.}(2023)\citenamefont{Tagami, Wakasa, and
  Yahiro}}]{Tagami}
\bibinfo{author}{\bibfnamefont{S.}~\bibnamefont{Tagami}},
  \bibinfo{author}{\bibfnamefont{T.}~\bibnamefont{Wakasa}}, \bibnamefont{and}
  \bibinfo{author}{\bibfnamefont{M.}~\bibnamefont{Yahiro}},
  \bibinfo{journal}{Results in Physics} \textbf{\bibinfo{volume}{46}},
  \bibinfo{pages}{106296} (\bibinfo{year}{2023}).

\bibitem[{\citenamefont{Fortson}(1993)}]{Fortson-Ion}
\bibinfo{author}{\bibfnamefont{N.}~\bibnamefont{Fortson}},
  \bibinfo{journal}{Phys. Rev. Lett.} \textbf{\bibinfo{volume}{70}},
  \bibinfo{pages}{2383} (\bibinfo{year}{1993}),
  \urlprefix\url{https://link.aps.org/doi/10.1103/PhysRevLett.70.2383}.

\bibitem[{\citenamefont{Zheng et~al.}(2026)\citenamefont{Zheng, Wang, Verma,
  Wang, Langin, and DeMille}}]{SnSimsPaper}
\bibinfo{author}{\bibfnamefont{G.}~\bibnamefont{Zheng}},
  \bibinfo{author}{\bibfnamefont{J.}~\bibnamefont{Wang}},
  \bibinfo{author}{\bibfnamefont{M.}~\bibnamefont{Verma}},
  \bibinfo{author}{\bibfnamefont{Q.}~\bibnamefont{Wang}},
  \bibinfo{author}{\bibfnamefont{T.~K.} \bibnamefont{Langin}},
  \bibnamefont{and} \bibinfo{author}{\bibfnamefont{D.}~\bibnamefont{DeMille}},
  \bibinfo{journal}{Phys. Rev. A} \textbf{\bibinfo{volume}{113}},
  \bibinfo{pages}{043115} (\bibinfo{year}{2026}),
  \urlprefix\url{https://link.aps.org/doi/10.1103/3g5j-f935}.

\bibitem[{\citenamefont{Dzuba}(2005)}]{Dzu05}
\bibinfo{author}{\bibfnamefont{V.~A.} \bibnamefont{Dzuba}},
  \bibinfo{journal}{Phys. Rev. A} \textbf{\bibinfo{volume}{71}},
  \bibinfo{pages}{032512} (\bibinfo{year}{2005}).

\bibitem[{\citenamefont{Johnson and Sapirstein}(1986)}]{B-splines}
\bibinfo{author}{\bibfnamefont{W.~R.} \bibnamefont{Johnson}} \bibnamefont{and}
  \bibinfo{author}{\bibfnamefont{J.}~\bibnamefont{Sapirstein}},
  \bibinfo{journal}{Phys. Rev. Lett.} \textbf{\bibinfo{volume}{57}},
  \bibinfo{pages}{1126} (\bibinfo{year}{1986}).

\bibitem[{\citenamefont{Dzuba}(2014)}]{Dzu-CI-SD14}
\bibinfo{author}{\bibfnamefont{V.~A.} \bibnamefont{Dzuba}},
  \bibinfo{journal}{Phys. Rev. A} \textbf{\bibinfo{volume}{90}},
  \bibinfo{pages}{012517} (\bibinfo{year}{2014}).

\bibitem[{\citenamefont{Dzuba et~al.}(2006)\citenamefont{Dzuba, Flambaum, and
  Safronova}}]{DzuFlaSaf06}
\bibinfo{author}{\bibfnamefont{V.~A.} \bibnamefont{Dzuba}},
  \bibinfo{author}{\bibfnamefont{V.~V.} \bibnamefont{Flambaum}},
  \bibnamefont{and} \bibinfo{author}{\bibfnamefont{M.~S.}
  \bibnamefont{Safronova}}, \bibinfo{journal}{Phys. Rev. A}
  \textbf{\bibinfo{volume}{73}}, \bibinfo{pages}{022112}
  (\bibinfo{year}{2006}).

\bibitem[{\citenamefont{Kramida et~al.}()\citenamefont{Kramida,
  {Yu.~Ralchenko}, Reader, and {NIST ASD Team}}}]{NIST}
\bibinfo{author}{\bibfnamefont{A.}~\bibnamefont{Kramida}},
  \bibinfo{author}{\bibnamefont{{Yu.~Ralchenko}}},
  \bibinfo{author}{\bibfnamefont{J.}~\bibnamefont{Reader}}, \bibnamefont{and}
  \bibinfo{author}{\bibnamefont{{NIST ASD Team}}}, \emph{\bibinfo{title}{{NIST
  Atomic Spectra Database \textnormal{(ver. 5.12), [Online]. Available: {\tt
  \url{https://physics.nist.gov/asd}} [2024, December 3]. National Institute of
  Standards and Technology, Gaithersburg, MD.}}}}, \bibinfo{note}{dOI:
  \href{https://doi.org/10.18434/T4W30F}{\tt https://doi.org/10.18434/T4W30F}.
  (2024).}

\bibitem[{\citenamefont{Dzuba et~al.}(2017{\natexlab{a}})\citenamefont{Dzuba,
  Berengut, Harabati, and Flambaum}}]{CIPT}
\bibinfo{author}{\bibfnamefont{V.~A.} \bibnamefont{Dzuba}},
  \bibinfo{author}{\bibfnamefont{J.~C.} \bibnamefont{Berengut}},
  \bibinfo{author}{\bibfnamefont{C.}~\bibnamefont{Harabati}}, \bibnamefont{and}
  \bibinfo{author}{\bibfnamefont{V.~V.} \bibnamefont{Flambaum}},
  \bibinfo{journal}{Phys. Rev. A} \textbf{\bibinfo{volume}{95}},
  \bibinfo{pages}{012503} (\bibinfo{year}{2017}{\natexlab{a}}).

\bibitem[{\citenamefont{Dalgarno and Lewis}(1955)}]{dalgarno1955exact}
\bibinfo{author}{\bibfnamefont{A.}~\bibnamefont{Dalgarno}} \bibnamefont{and}
  \bibinfo{author}{\bibfnamefont{J.~T.} \bibnamefont{Lewis}},
  \bibinfo{journal}{Proc. R. Soc. A} \textbf{\bibinfo{volume}{233}},
  \bibinfo{pages}{70} (\bibinfo{year}{1955}).

\bibitem[{\citenamefont{Dzuba et~al.}(1987)\citenamefont{Dzuba, Flambaum,
  Silvestrov, and Sushkov}}]{DzuFlaSilSus87}
\bibinfo{author}{\bibfnamefont{V.~A.} \bibnamefont{Dzuba}},
  \bibinfo{author}{\bibfnamefont{V.~V.} \bibnamefont{Flambaum}},
  \bibinfo{author}{\bibfnamefont{P.~G.} \bibnamefont{Silvestrov}},
  \bibnamefont{and} \bibinfo{author}{\bibfnamefont{O.~P.}
  \bibnamefont{Sushkov}}, \bibinfo{journal}{J. Phys. B}
  \textbf{\bibinfo{volume}{20}}, \bibinfo{pages}{1399} (\bibinfo{year}{1987}).

\bibitem[{\citenamefont{Dzuba}(2020)}]{Symmetry}
\bibinfo{author}{\bibfnamefont{V.}~\bibnamefont{Dzuba}},
  \bibinfo{journal}{Symmetry} \textbf{\bibinfo{volume}{12}},
  \bibinfo{pages}{1950} (\bibinfo{year}{2020}).

\bibitem[{\citenamefont{Dzuba et~al.}(1988)\citenamefont{Dzuba, Flambaum,
  Silvestrov, and Sushkov}}]{DzuFlaSilSus88}
\bibinfo{author}{\bibfnamefont{V.~A.} \bibnamefont{Dzuba}},
  \bibinfo{author}{\bibfnamefont{V.~V.} \bibnamefont{Flambaum}},
  \bibinfo{author}{\bibfnamefont{P.~G.} \bibnamefont{Silvestrov}},
  \bibnamefont{and} \bibinfo{author}{\bibfnamefont{O.~P.}
  \bibnamefont{Sushkov}}, \bibinfo{journal}{Europhys. Lett.}
  \textbf{\bibinfo{volume}{7}}, \bibinfo{pages}{413} (\bibinfo{year}{1988}).

\bibitem[{\citenamefont{Porsev et~al.}(2016)\citenamefont{Porsev, Kozlov,
  Safronova, and Tupitsyn}}]{Safronova}
\bibinfo{author}{\bibfnamefont{S.~G.} \bibnamefont{Porsev}},
  \bibinfo{author}{\bibfnamefont{M.~G.} \bibnamefont{Kozlov}},
  \bibinfo{author}{\bibfnamefont{M.~S.} \bibnamefont{Safronova}},
  \bibnamefont{and} \bibinfo{author}{\bibfnamefont{I.~I.}
  \bibnamefont{Tupitsyn}}, \bibinfo{journal}{Phys. Rev. A}
  \textbf{\bibinfo{volume}{93}}, \bibinfo{pages}{012501}
  (\bibinfo{year}{2016}).

\bibitem[{\citenamefont{Meekhof et~al.}(1993)\citenamefont{Meekhof, Vetter,
  Majumder, Lamoreaux, and Fortson}}]{Pb-PNC1}
\bibinfo{author}{\bibfnamefont{D.~M.} \bibnamefont{Meekhof}},
  \bibinfo{author}{\bibfnamefont{P.~A.} \bibnamefont{Vetter}},
  \bibinfo{author}{\bibfnamefont{P.~K.} \bibnamefont{Majumder}},
  \bibinfo{author}{\bibfnamefont{S.~K.} \bibnamefont{Lamoreaux}},
  \bibnamefont{and} \bibinfo{author}{\bibfnamefont{E.~N.}
  \bibnamefont{Fortson}}, \bibinfo{journal}{Phys. Rev. Lett.}
  \textbf{\bibinfo{volume}{71}}, \bibinfo{pages}{3442} (\bibinfo{year}{1993}).

\bibitem[{\citenamefont{Meekhof et~al.}(1995)\citenamefont{Meekhof, Vetter,
  Majumder, Lamoreaux, and Fortson}}]{Pb-PNC2}
\bibinfo{author}{\bibfnamefont{D.~M.} \bibnamefont{Meekhof}},
  \bibinfo{author}{\bibfnamefont{P.~A.} \bibnamefont{Vetter}},
  \bibinfo{author}{\bibfnamefont{P.~K.} \bibnamefont{Majumder}},
  \bibinfo{author}{\bibfnamefont{S.~K.} \bibnamefont{Lamoreaux}},
  \bibnamefont{and} \bibinfo{author}{\bibfnamefont{E.~N.}
  \bibnamefont{Fortson}}, \bibinfo{journal}{Phys. Rev. A}
  \textbf{\bibinfo{volume}{52}}, \bibinfo{pages}{1895} (\bibinfo{year}{1995}).

\bibitem[{\citenamefont{Phipp et~al.}(1996)\citenamefont{Phipp, Edwards, Baird,
  and Nakayama}}]{Pb-PNC3}
\bibinfo{author}{\bibfnamefont{S.~J.} \bibnamefont{Phipp}},
  \bibinfo{author}{\bibfnamefont{N.~H.} \bibnamefont{Edwards}},
  \bibinfo{author}{\bibfnamefont{P.~E.~G.} \bibnamefont{Baird}},
  \bibnamefont{and} \bibinfo{author}{\bibfnamefont{S.}~\bibnamefont{Nakayama}},
  \bibinfo{journal}{J. Phys. B: At. Mol. Opt. Phys.}
  \textbf{\bibinfo{volume}{29}}, \bibinfo{pages}{1861} (\bibinfo{year}{1996}).

\bibitem[{\citenamefont{Schwerdtfeger and Nagle}(2019)}]{PolTab}
\bibinfo{author}{\bibfnamefont{P.}~\bibnamefont{Schwerdtfeger}}
  \bibnamefont{and} \bibinfo{author}{\bibfnamefont{J.~K.} \bibnamefont{Nagle}},
  \bibinfo{journal}{Mol Phys.} \textbf{\bibinfo{volume}{117}},
  \bibinfo{pages}{1200} (\bibinfo{year}{2019}).

\bibitem[{\citenamefont{Dzuba et~al.}(2017{\natexlab{b}})\citenamefont{Dzuba,
  Flambaum, and Stadnik}}]{DFS17}
\bibinfo{author}{\bibfnamefont{V.~A.} \bibnamefont{Dzuba}},
  \bibinfo{author}{\bibfnamefont{V.~V.} \bibnamefont{Flambaum}},
  \bibnamefont{and} \bibinfo{author}{\bibfnamefont{Y.~V.}
  \bibnamefont{Stadnik}}, \bibinfo{journal}{Phys. Rev. Lett.}
  \textbf{\bibinfo{volume}{119}}, \bibinfo{pages}{223201}
  (\bibinfo{year}{2017}{\natexlab{b}}).

\bibitem[{\citenamefont{Dzuba et~al.}(2026)\citenamefont{Dzuba, Flambaum, and
  Vong}}]{DFV26}
\bibinfo{author}{\bibfnamefont{V.~A.} \bibnamefont{Dzuba}},
  \bibinfo{author}{\bibfnamefont{V.~V.} \bibnamefont{Flambaum}},
  \bibnamefont{and} \bibinfo{author}{\bibfnamefont{G.~K.} \bibnamefont{Vong}},
  \bibinfo{journal}{Phys. Rev. A} \textbf{\bibinfo{volume}{113}},
  \bibinfo{pages}{052808} (\bibinfo{year}{2026}).

\bibitem[{\citenamefont{Angeli and Marinova}(2013)}]{Marinova}
\bibinfo{author}{\bibfnamefont{I.}~\bibnamefont{Angeli}} \bibnamefont{and}
  \bibinfo{author}{\bibfnamefont{K.}~\bibnamefont{Marinova}},
  \bibinfo{journal}{Atomic Data and Nuclear Data Tables}
  \textbf{\bibinfo{volume}{99}}, \bibinfo{pages}{69} (\bibinfo{year}{2013}).

\bibitem[{\citenamefont{Antypas and Elliott}(2013)}]{Elliot-PNCScheme}
\bibinfo{author}{\bibfnamefont{D.}~\bibnamefont{Antypas}} \bibnamefont{and}
  \bibinfo{author}{\bibfnamefont{D.~S.} \bibnamefont{Elliott}},
  \bibinfo{journal}{Phys. Rev. A} \textbf{\bibinfo{volume}{87}},
  \bibinfo{pages}{042505} (\bibinfo{year}{2013}),
  \urlprefix\url{https://link.aps.org/doi/10.1103/PhysRevA.87.042505}.

\bibitem[{\citenamefont{DeMille et~al.}(1999)\citenamefont{DeMille, Budker,
  Derr, and Deveney}}]{demille1999search}
\bibinfo{author}{\bibfnamefont{D.}~\bibnamefont{DeMille}},
  \bibinfo{author}{\bibfnamefont{D.}~\bibnamefont{Budker}},
  \bibinfo{author}{\bibfnamefont{N.}~\bibnamefont{Derr}}, \bibnamefont{and}
  \bibinfo{author}{\bibfnamefont{E.}~\bibnamefont{Deveney}},
  \bibinfo{journal}{Phys. Rev. Lett.} \textbf{\bibinfo{volume}{83}},
  \bibinfo{pages}{3978} (\bibinfo{year}{1999}).

\bibitem[{\citenamefont{Takahashi and {\em et al.}}(2026)}]{SM2026}
\bibinfo{author}{\bibfnamefont{F.}~\bibnamefont{Takahashi}} \bibnamefont{and}
  \bibinfo{author}{\bibnamefont{{\em et al.}}}
  (\bibinfo{collaboration}{Particle Data Group}), \bibinfo{journal}{Int. J.
  Mod. Phys. A} \textbf{\bibinfo{volume}{41}}, \bibinfo{pages}{2630011}
  (\bibinfo{year}{2026}).

\end{thebibliography}

\end{document}